\documentclass{article}
\usepackage{graphicx} 
\usepackage{authblk}
\usepackage{url}
\usepackage{booktabs}
\usepackage{geometry}

\newgeometry{vmargin={2cm}, hmargin={2.5cm, 2.5cm}}

\begin{document}

\title{Understanding \#vent Channels on Discord}
\author{Kayode Oladeji}
\affil{Georgia Institute of Technology}

\author{Tony Wang\footnote{Work done at the Georgia Institute of Technology.}}
\affil{Cornell University}

\author{Diyi Yang}
\affil{Stanford University}

\author{Amy Bruckman}
\affil{Georgia Institute of Technology}

\date{September 2024}

\maketitle

\section{Abstract}

Vent channels on Discord, which are chat channels developed for people to express frustrations, can become an informal type of peer support system. This paper is a qualitative study of experiences with vent channels on Discord, examining the experiences of 13 participants through semi-structured interviews. We find that participants are able to meet their needs for social support via vent channels by receiving commiseration, advice, and validation from the responses of others. At the same time, vent channels can lead to frustration when participants have conflicting expectations for their interactions. We suggest ways that Discord or Discord server moderators can provide enhanced structure, clarity, and transparency in order to enable participants to have better experiences in vent channels. 

\section{Introduction}
Self-disclosure is a key mechanism for receiving emotional and informational support in online communities \cite{Luo2020-sp, Yang2019-kz}, but the posting of content with emotions such as anger or frustration can negatively impact community members who view such content. Self-disclosure of negative emotions, or venting, occurs online in various platforms such as Reddit \cite{Balani2015-ji}, Instagram \cite{Andalibi2017-om}, and Twitter \cite{Vermeulen2018-xv}. However, few of those spaces appear to have the high levels of responsive interaction and support seen in Discord channels, a platform that has seen significant growth in recent years particularly among adolescents and young adults who have traditionally been a target of study for online self-disclosure research.

The proposed research focuses on understanding the mechanism of online “vent channels” hosted on Discord. Discord is a messaging platform launched in 2015 that allows for communication in community-based hubs, known as “servers” \cite{noauthor_undated-zx}. It allows content to be organized by topic-related channels within an individual server, and it provides voice channels which allow users to communicate via audio and video. Vent channels are generally described within Discord as spaces for community members to vent about stressful events in daily life and receive support from other members.

In vent channels, responders typically provide social support to venters by empathizing and providing advice, which helps venters to feel comforted and less isolated. Participants use venting channels as spaces to self-disclose negative emotions and receive support from others for many reasons, such as anonymity or because they find general community and social support on Discord. An example of a venter-responder exchange on a vent channel may resemble the following: 

\begin{quote}
Venter: "Ugh, I'm sick today and I have to give a presentation. I'm so frustrated" 

Responder: "That sucks, I hope you get better soon!"
 \end{quote}

Vent channels are a popular feature on Discord. At the time of writing, the Lofi Girl server, which has over 800,000 members and is the 12th largest Discord server \cite{noauthor_undated-vb} has two vent channels which are visible to any member of the server. The Study Together server, with over 500,000 members, also has a vent channel. Upon searching for “vent” or “mental health” under Discord’s Server Discovery tab, a way to browse publicly available servers, there are dozens of servers with the express purpose of serving as a place for people to vent about their frustrations.

In this study, we aim to gain understanding of the participants of vent channels and their
experiences, both positive and negative. We focus this work around the following research questions: 

\begin{enumerate}
    \item What are the experiences of people who vent on vent channels? 
    \item What are the experiences of people who respond to others' vents?
    \item What do participants want to get out of vent channels, and do they receive it? 
    \item How could vent channels be redesigned to better meet participants' needs?
\end{enumerate}

\section{Related Work}
Social psychologist Jennifer Parlamis discusses the action of venting as follows:  

\begin{quote}
    Venting of emotions has been described, metaphorically, as similar to venting a steamfilled pipe. If you allow the steam in a pipe to escape by venting it, the pressure will be released and the pipe will not explode. Similarly, if an individual vents emotions by expressing them, the emotions will dissipate and the individual will return to a more placid state. \cite{Parlamis2012-qv} \end{quote} 

The common popular perception, which adheres to Parlamis’ description, is that venting out negative emotions such as anger or sadness by expressing them to others has the positive effect of reducing these negative emotions for the person experiencing them \cite{Zech2005-ql}. However, Nils and Rimé offer research that provides evidence against the venting hypothesis: in their study, participants who verbally vented about an aversive video were more likely to continue to experience negative emotions about it \cite{Nils2012-tt}. This result does come with a significant caveat, which is that if the responder engaged in cognitive work with the person venting, emotional recovery is more likely. Cognitive work includes strategies such as getting the venter to view the situation objectively and generally aiming to increase the venter’s sense of agency. While not always the case, much of the time other participants in a vent channel will respond to a vent and converse with the venter in ways that may emulate this process of cognitive work.  

We are also interested in understanding the effects of venting on those who read and engage with the vents of others. In “Anger on the Internet,” Martin et. al examine online sites for ranting, which are a similar phenomenon as channels for venting, but with a particular focus on anger \cite{Martin2013-hk}. They found that when participants were asked to behave like a typical member on the studied site, they became less happy and more sad or angry. The authors state that “reading and writing online rants are associated with negative shifts in mood for the vast majority of people,” which is especially relevant to our study given that vent channels on Discord include rants or vents that focus on anger and frustrating situations.  

In a related work, Christine Elliot Chin’s thesis  studies the r/2meirl4meirl Discord server and its included vent channel \cite{Chin2021-bn}. Her assessment of the vent channel in this particular community is as an “anti-community” space, where users do not interact with each other but merely vent in the same place. The vent channel is seen as a void, and content posted there is not discussed or continued in other channels of the server. She observes that despite the lack of direct interaction, the vent channel serves as a method of validating participants’ experiences and forming unstated bonds between them. Members who have a mental illness can relate to and feel relieved by the vents of other members with similar conditions and struggles.  

Much of what is interesting about vent channels is in the self-disclosure of users. In many situations, members who participate in vent channels may not know or have interacted with one another at all, and yet they continue to divulge highly personal and sensitive information to an audience that is unknown and often unrestricted. The psychological concept of self-disclosure explains much of why people participate in this behavior. In “The Channel Matters," Yang et. al research the act of self-disclosure in online cancer support groups \cite{Yang2019-kz}. They find that in the act of self-disclosure, participants weigh the potential costs, such as loss of anonymity and loss of control over information, against the potential benefits, such as emotional support, help, and advice. A very similar thing occurs in Discord vent channels, where the expected potential benefits to participants are greater than any expected harm from loss of privacy. One differentiating factor in the context of this paper is that vent channels are not typically focused on a particular subject of common experience like the cancer support groups, but rather, in most cases, are left open-ended to whatever topics participants want to discuss.  

Finally, in “Self-disclosure and Social Media”, Luo and Hancock synthesize research to develop a framework of the relationships between self-disclosure and mental wellbeing, resulting in four primary mechanisms by which the former can affect the latter \cite{Luo2020-sp}. These are perceived connectedness (perceiving a connection with other people), social support (the social benefits recieved from social interactions), capitalization process (feeling happy about sharing positive information and sad about sharing negative information), and perceived authenticity (perceiving that one is able to present oneself in an authentic way). For example, some people self-disclose with the goal of feeling connected to others and have increased mental well-being when they are able to achieve this through engaging with others who have similar situations or difficulties. Others may have motivations of validation and will experience increased mental well-being when this validation is given by other users in response to a vent. The impact of self-disclosure on mental wellbeing is highly dependent on individual motivations and how well they are met, which in itself can be affected by the duration of disclosure, honesty of disclosure, self-esteem of the discloser, and engagement of others who interact with the discloser.  

\section{Methods}

This study aims to increase understanding of experiences with channels for venting on Discord, why participants use vent channels, and what benefits they receive. To gain this information, we conducted 13 semi-structured interviews with vent channel participants, supporters, and moderators to understand their experiences.

\subsection{Participants}
To recruit participants, we used a combination of convenience sampling and snowball sampling. The first author advertised this study in Discord servers in which he was a pre-existing member, which included a university community, a casual music community, and a marginalized-identity-specific community. In addition, we recruited through several Discord-related subreddits on reddit.com, such as r/discordapp and r/DiscordAdvertising. Servers used in recruitment are described in Table \ref{tab:servers}. Participant recruitment occurred from November 2022 to April 2023 and was stopped when we reached saturation and began to hear repeated themes in interviews.

\begin{table}[htp]\centering
\caption{Discord servers used in recruitment}
\label{tab:servers}
\begin{tabular}{ccc}
    \toprule
    Server Topic & Server Size & Participants \\
    \midrule
    Academic & 1000+ & P1, P3, P9, P11 \\
    Music & 500-1000 & P2 \\
    Identity & 100-500 & P5, P6, P8 \\
    Fandom & 100-500 & P7 \\
    Venting & 1000+ & P12 \\
\bottomrule
\end{tabular}
\end{table}

Participants had a range of roles within their Discord servers –- we had representation from users who used vent channels purely to vent and did not respond to others, users who only responded and did not vent themselves, users who did both, and users who moderated the vent channel. Participant demographics are described in detail in Table \ref{tab:demo}.  

\begin{table*}[htp]\centering
\caption{Participant demographics}
\label{tab:demo}
\begin{tabular}{ccccccll}
    \toprule
    Participant&Venter&Responder&Moderator&Time on Discord & Age & Gender & Race/Ethnicity\\
    \midrule
P1  & X                          & X                             & X                             & 6 years                             & 20                      & Man                        & White                              \\
P2  &                            & X                             &                               & 6 years                             & 23                      & Nonbinary                  & South Asian                        \\
P3  & X                          &                               & X                             & 5 years                             & 23                      & Woman                      & East Asian                         \\
P4  &                            &                               & X                             & 5 years                             & 31                      & Man                        & White                              \\
P5  & X                          & X                             & X                             & 6 years                             & 26                      & Man                        & Black                              \\
P6  & X                          & X                             &                               & 4 years                             & 24                      & Nonbinary                  & Mixed                              \\
P7  &                            &                               & X                             & 2 years                             & 27                      & Woman                      & White                              \\
P8  & X                          &                               &                               & 3 years                             & 21                      & Nonbinary                  & Black                              \\
P9  & X                          & X                             &                               & 6 years                             & 21                      & Man                        & Hispanic                           \\
P10 & X                          & X                             &                               & 6 years                             & 24                      & Nonbinary                  & Mixed                              \\
P11 & X                          & X                             &                               & 5 years                             & 19                      & Man                        & Hispanic                           \\
P12 & X                          & X                             & X                             & 4 years                             & 18                      & Man                        & Mixed                              \\
P13 & X                          & X                             &                               & 6 years                             & 18                      & Man                        & White   \\
\bottomrule
\end{tabular}
\end{table*}

\subsection{Interview Procedure}
During interviews, the researchers aimed to learn about participants’ experiences with venting, responding to others, and moderating in vent channels. The full text of interview questions is listed in the supplemental material \ref{interview-questions}. Interview questions were grouped into six major categories:  

\begin{enumerate}
    \item \textbf{Introductory questions.} These questions were used in order to gain context for the participant’s experiences with vent channels. How long has the participant used Discord, what sorts of communities do they frequent, and what is their experience with the server in question outside of the vent channel?  

    \item \textbf{Experience with venting.} These questions centered around the act of posting content in the vent channel. How often do you participate in the vent channel, and what do you typically do? What do you share when you vent, and what do you want to happen as a result? Do you worry about privacy when venting?  

    \item \textbf{Experience with supporting.} These questions focused on the act of responding to post from others in the vent channel. Can you tell me about a time when you responded to someone else on the vent channel? How do you feel after spending time on the vent channel? Has anyone ever worried you because of something they posted?  

    \item \textbf{Experience as a moderator.} These questions were asked about the experience of moderators on the vent channel. What sort of content typically needs to be moderated? Does the vent channel have rules, and if so, how were those crafted?  

    \item \textbf{Introspective.} These questions examined participants’ opinions on vent channels and their functionality. How would you describe the purpose of a vent channel? What features have improved them, in your experience?  

    \item \textbf{Demographics.} Participants were asked their age, gender identity, pronouns, race and/or ethnicity, and country of residence.  
\end{enumerate}

These questions were used as a starting point for the interviewer, who would then flexibly follow interesting lines of inquiry during the interview. Eight interviews were conducted over audio, one was conducted in-person, and four were conducted over synchronous text. All interviews were limited to a time of 90 minutes, and audio and synchronous text interviews took place over Discord direct messaging. The audio and in-person interviews were recorded, transcribed using the automated service otter.ai, and finally edited by the authors for correctness prior to analysis. 

\subsection{Analysis}
Authors performed a thematic analysis of the interview transcripts, using inductive open coding based on Braun and Clark’s method \cite{Braun2006-if}. These codes were then organized into higher-level themes, which revealed the findings discussed below. Transcripts were coded individually by the lead author with an initial set of themes was derived after five interviews. These themes were then reviewed with other authors before initiating additional reflection and refinement of codes and themes. This process continued as interviews were conducted until P11 and P13's interviews did not provide new information, which suggested saturation in our themes and enough data had been collected.

\subsection{Ethical Considerations}
This study was approved by the primary institution’s IRB and was deemed a low level of risk. In accordance, a plan was developed for interacting with participants who disclosed suicidal ideation, which included immediately ceasing interviewing, expressing concern for their wellbeing, and referring them to mental health hotlines and other professionals. This did not occur during our interviews.

In terms of relevant identities and positionality, the first author is a twenty-year-old recent alumnus of an undergraduate program. He first used vent channels four years ago in a Discord server primarily consisting of, and run by, other students. For several months, he moderated a vent channel aimed at people undergoing life transitions. This experience sparked his interest in online communities. He has used and observed vent channels on Discord for four years, in a variety of contexts. The second, third, and fourth authors have not used vent channels regularly in the past. 

\section{Findings}

Next, we discuss our findings from participant interviews and their relevance to our research questions. We first discuss the overarching themes regarding the purpose of vent channels, the ways that they are used, and how participants feel about them.  

\subsection{About Vent Channels} 
\subsubsection{Purpose and Use}

A vent channel is generally portrayed as a space for users to express negative emotions and frustrations. Two of the channels studied had the following descriptions as labels for users to understand the type of content that would fit there: ``vent about anything you need to get off your chest," and ``for venting your issues and emotions." While this description may be the primary purpose of the vent channel, it is rarely the sole outcome. While some moderators described the channel as a space to express emotions, non-moderating participants frequently added a description of the channel as being a place to receive support. P5 described them as “a place where people can come to vent their frustrations and ask for advice if they need … or just generally seek support when they're going through some rough times.” P13 emphasized that participating in a vent channel comes with an “expectation of support,” further elaborating that positivity and support should be a channel norm. He detailed: 

\begin{quote}    
 IMO, the fact that it's a vent channel in and of itself means that you should be able to use the channel to talk about whatever is troubling you or whatever is going on in your life, and then expect nothing but support in response. If people are rude to you in a vent channel, the entire purpose of the channel is defeated. You might as well delete it.  \end{quote}

At times, the channel norm of providing a response can cause conflict with the goals of the venter, which may simply be to vent frustration without extended interaction. We will discuss this distinction further in section \ref{responses-support}.  

While the participants in this study ranged from ages 18 to 31, during recruitment, the primary author discovered multiple Discord servers with vent channels where the majority of the participants were under 18 (based on self-selected user tags). Additionally, P12 stated that the majority of participants in the venting-centered Discord server he moderates are under 18. Due to the difficulty of gaining the necessary parental consent, we opted not to study participants under 18, but we want to note the prevalence of minors in vent channels.  

Participants described using vent channels for two major reasons: to get things off their chest in a similar means as shouting into a void, and to ask others for advice while receiving external perspectives on their situation.  For example, P1 described using a vent channel to inform members of a community about a negative event in his life. P6 described using a vent channel in a server for members of a marginalized group as a topic-specific support group where he could vent about specific struggles as a marginalized person to others who would understand and empathize. Finally, P4 described witnessing participants use the vent channel that he moderates as a way to receive attention and validation. 
 
 \subsubsection{Vent Channel Structure}
 In most of the channels studied, a vent channel is just one of many topic-focused text channels within a larger Discord server. Vent channels appear in a variety of Discord servers, not only those focused on mental health. For example, three servers studied in this project include a server for students of a post-secondary university, a server for fans of a music artist, and a server for members of a marginalized identity group.  

However, we interviewed a moderator of one server that branded itself as a “venting server” and contained over forty channels focused on venting of some kind. These channels were largely topic-dependent (e.g. relationship issues, LGBTQIA+ identity struggles, or anxiety) as well as specifying the form of venting. One channel in this server serves specifically as a place to ask for advice, while another channel centers around venting through art.  

The two primary roles observed in vent channels are venter and responder. A venter is a participant who makes a post to the vent channel expressing a negative emotion or frustration, while a responder is a participant who responds to a specific vent. While some participants interviewed described only participating in one of these two roles, the roles frequently overlap. A person who vents on one day could respond to the vents of others another day. The typical vent channel is very much a peer-supported environment with no traditional hierarchy of those who help and those who receive help. However, in one server centered around mental health support, these roles were more structured, with users able to select a “role,” a way of grouping users on Discord, indicating that they would like to provide support. In this situation, when a person vents, they’re able to ping this role, so that every user who has selected the role is notified that someone would like support.  

Few vent channels defined any sort of limit on the severity of emotion that could be expressed in a channel. As a result, the content posted in a vent channel varies drastically from light frustrations such as “I hate that every song at karaoke is out of my vocal range,” to more severe existential concerns such as “I feel trapped by where I am in life and progress feels unattainable,” to even discussion of suicidality.  

A final role in a vent channel is that of a moderator, which we will discuss in more detail later. Moderators make decisions about what sort of content and behavior, if any, should be banned or removed from the vent channel. In situations where the severity of a user’s vent escalates above the point where the vent channel could provide meaningful help, this removal is typically seen as the responsibility of the moderator.  

\subsection{User Experiences with Self-Disclosure}
\label{user-experiences-use}

We found that vent channels had both social and cognitive functions for community members on Discord. Their use fosters perceptions of connectedness, authenticity, and serve as a way to receive anonymous advice similar to other online health communities \cite{Yang2019-kz}. Participants described feeling connected to other people who used the vent channel because they learned more about others when they vented. P6 described how seeing activity in vent channels impacts how connected they felt to other members: 
\begin{quote} 
    When I see [other users] posting in the venting channel, it gives me a little bit more context and background for their real life lives and who they are as a person outside of online spaces … It's cool to get a deeper understanding of who they are as a person based on what they are choosing to share that they struggle about. 
\end{quote}
Other participants noted that, even though it is unpleasant to see other people venting about sad things happening to them, it can be nice to see familiar usernames and feel a sense of community. P2 suggested that vent channels are more for providing connection and building community than for providing individual support, saying:  
\begin{quote}
    I think the core of venting is not talking about your problems and feeling better. The core of vent channels purposefully [sic] to me is like, be vulnerable and share yourself with other people and feel connected. 
\end{quote}

Much of the interaction in vent channels relates to the second mechanism, social support. Essentially, people receive social support from being noticed and being supported, and the vent channel is first and foremost a venue for participants to be heard. All the vent channels studied were visible by other users and intended to be viewed by other users, and at times, just the knowledge that their message has been seen by another person can be helpful for a venter. In addition to this dynamic, many vent channels feature frequent responses to vents, which can express social support in the form of reassurance, validation, or sympathy. Communal approaches to fostering perceived connectiveness among members can impact members' willingness to self-disclose \cite{Luo2020-sp}, creating a mechanism for participation within a community.  

Furthermore, when venters share vulnerable situations and feelings with others, it can increase feelings of authenticity as well as increase their connection to the community. P2 described that “vulnerability is what makes a relationship,” and that they believe that vent channels can facilitate community because the authenticity of venters fosters connections. This notion of authenticity through the expression of negative emotions is a consistent phenomenon aligned with prior work \cite{Luo2020-sp}.

Finally, the act of formulating messages about a frustrating situation or difficult feeling could help venters to process and understand what they’re feeling, and communicating with other people about it could further help. As an extension of this idea, some participants mentioned using vent channels to get feedback on an external situation, processing both their thoughts about it and reaction to it with people who have an outside perspective as a way of seeing if their thoughts and reactions are reasonable.  

\subsection{Responses and Support}
\label{responses-support}
There are a few groups of responses common to vent channels: commiseration, validation, reassurance, and sympathy. Other users will commiserate and validate the frustration of a venter (e.g. “I know the feeling, it really sucks”) or offer reassurance (e.g. “It’ll be ok! I know you can do this”). The most general type of response is expressing sympathy along the lines of “That really sucks, and I’m sorry it’s happening to you.”  

Some users gain satisfaction from helping others in vent channels, so much so that they intentionally take on the role and will volunteer to give advice and support. P2 is a member of a Discord server that is centered around emotional support and has said that they “really enjoy helping people in that way.” Many participants expressed feelings of gratification when they are able to successfully help another person, describing it as a fulfilling, feel-good moment. In tandem, however, participants also felt overwhelmed or frustrated by giving support, especially when venters were unreceptive. P3 used to try to help venters, but eventually, the activity “started to become a little bit of an energy sucker.” She stated that trying to talk someone out of negative thoughts requires strategy and tact. P4 also no longer responds to vents, because he feels that watching vent channels is like “volunteering to be a therapist full-time.”  

A distinct category of response includes problem-solving and offering advice. Sometimes other respondents seek a way to make the venter’s situation better, asking “What can I do to help? How can your difficulty be fixed?” While some users come to a vent channel specifically looking for advice, many do not, and if this intention is not clarified, it can lead to misunderstandings and negative interactions. This categorization is a sensitive area. P5 stated that as a moderator, he has had to deal with levels of miscommunication, where supporters in a vent channel will aim to “answer a question that wasn’t really asked” or solve problems without any indication that problem-solving is what the venter desires. P3 stated that when she posted in the vent channel with frustrations about being sick, people would respond, “Are you okay? Have you gotten any medication or medical attention for your sickness? Why are you sick?” This sort of inquiry could be helpful for some, interpreted as merely expressing support, but some users just want to vent, without other users trying to help them solve problems. P5 believed that this confusion could be a negative experience, saying that “if people are coming to this channel, they're already probably not in the best emotional state. And many people do not respond well, including myself, to unsolicited advice.” This advice can also make venters feel like they are not being heard or understood because they receive advice for which they did not ask, and possibly because they do not receive the form of response that they desired.  

\begin{figure}[htp]
  \centering
  \includegraphics[width=\linewidth]{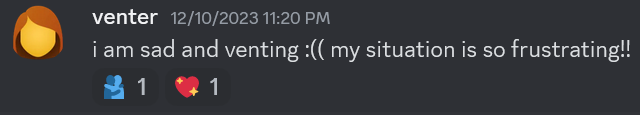}
  \caption{Appearance of a Discord Emoji Reaction}
  \label{reaction-pic}
\end{figure}

One of Discord’s built-in features is the “reaction,” where users are able to append an emoji to a message (Figure \ref{reaction-pic}), which is often meant to represent an emotional response to the message. Multiple participants noted that receiving supportive reactions to their message, for example heart emojis or hug emojis, made them feel heard. Participants described using these forms of acknowledgement to support others when they were “low on energy,” (P13) unsure of exactly what to say to the venter, or when they felt that the venter may not want to be engaged with in terms of textual conversation. Reacting with a positive emoji is a low-stakes and generally inoffensive gesture, which aligns with it being common, occurring at least once in every channel studied. By reacting with a supportive emoji, responders are able to avoid overstepping when a full textual response may not be desired by the venter, while still providing some form of support and acknowledgement.  

While much of the time other users will respond to and interact with venters, sometimes messages that resemble cries for help go unanswered. P3 described people who come to the vent channel and post statements like “I don’t know if I’ll ever be loved" that can go unanswered by other participants. Multiple participants described a lack of response as potentially being caused by a bystander effect \cite{guazzini2019online}, where readers of the channel had the feeling that someone else would respond. Participants had described avoided responding because they did not know what to say, or worried about making a venter feel worse. P11 described how a lack of response could affect others, saying that “the silence can feel deafening … They may end up feeling even worse than from when they initially posted the message or even lash out depending on the context.” When people post in vent channels, they are typically in some form of distress, and a lack of response could cause feelings of being neglected or abandoned, especially if the venter has an expectation of receiving support.   

\subsection{Positive Experiences}

What positive experiences do participants have with vent channels? Generally, venters feel positive about receiving the desired interaction after posting a vent, whether that’s others simply leaving a supportive reaction, giving detailed advice, or expressing sympathy. They feel supported and relieved. P6 described his feelings after posting to a vent channel saying, “I usually feel better after getting it off my chest. And then if I have people replying to me and responding to that, I definitely feel much better and comforted.”  

Participants also describe that the act of reading through the vent channel can help them to feel less alone. P3 stated that ``it helps with isolation; other people are going through the same things.” P6 described that as a member of a marginalized community and a Discord server centered around that community, the vent channel in that server helps him to know that other members of the community have shared experiences. As he said, "It’s really good to see that other people have gone through similar things that are out there, and that they’re making it through as well.” P3 described that it feels nice to see familiar usernames in the channel helping each other and makes the space feel like a community.  

Responders also described many positive feelings associated with being able to help a venter. When venters are receptive, respond positively to replies, or indicate that they feel better than at the original time of posting, responders feel fulfilled and happy that they have been able to provide help for someone else. P6 stated: 
\begin{quote}
    If people have responded positively to my reply to them, then that also usually feels pretty good. Because it's like, okay, cool, what I said did actually help this person in some way. And that's pretty meaningful.  
\end{quote}
While the process of supporting others can be difficult, many responders feel that the positive moments stand out above, or outweigh, the negative experiences. P13 stated about supporting others: “I've gotta be honest, it's pretty emotionally draining. I don't do it very often because it’ll burn me out if I’m not careful. But when I do and it makes them feel better, it's 100\% worth it.”  

\subsection{Negative Experiences}
We want to understand what negative experiences participants had with vent channels in order to mitigate them. One common experience was that vent channel could at times feel like a stream of negativity, which made it difficult to engage. Reading all of the struggles and difficulties of other members can be emotionally exhausting, especially when people post about intense situations involving harm. A vent channel can be dour and can bring the mood down. P10 described that they avoid reading the vent channel unless they are in a period of low stress and good mental health, because the vent channel can worsen their own mental health. Many participants described that vent channels could frequently be overwhelming in terms of the intensity of negative emotions. This intensity resulted in the participants avoiding reading the channel and spending less time responding to the vents of others. P13 stated, “it's draining just to read, let alone try to help,” and P10 echoed this distress, stating, “I do care, but it is so difficult and it's very heartbreaking. I wish I had the mental capacity to deal with all of it, but I'm only one human with one brain and I'm not trained in this.”  

In most interviews, the first author asked the question “Has anyone ever worried you because of what they posted in the vent channel?” and every participant who was asked this question responded yes. Sometimes the people who post to vent channels are experiencing a mental health crisis featuring thoughts of suicide or self-harm and will vent about these things, which can often leave other participants in the vent channel worried about the safety of this individual and concerned for their well-being.  

Situations like these are particularly concerning because people in the vent channel may not know any personal information about the person in crisis other than their Discord username. In addition, they are not mental health professionals, so as P4 stated, “It's very hard to give any sort of meaningful sound advice outside of ‘you need to reach out to this organization.’” Participants typically have no way to help the individual other than through words, and if the individual stops responding, they have no way of knowing what happened. This uncertainty can be stressful and scary.  

Moderators face many challenges relating to vent channels, but one of the major concerns is about their liability. Moderators tend to feel some level of responsibility for the content posted in their servers and events that happen as a result.  As P5 described,  
\begin{quote} As far as we know, nobody in the server is a therapist and we just can't necessarily take on the responsibility of therapizing people. So if there's talk of self-harm or things like that or threats to other people's lives, it's just a difficult thing to take on. \end{quote}
Multiple participants described that when a member posts content related to suicide, a moderator would message the member, asking about their well-being and providing resources such as crisis hotline phone numbers. 

Responders also described having frustrating experiences when they tried to help a venter and the person was unresponsive to their suggestions. P13 in particular described vent channels as sometimes fostering a culture of helplessness, where venters did not exhibit self-efficacy and seemed to stay stuck in the same difficulty situation, repeatedly venting and receiving sympathy rather than attempting to improve their lives.   

\subsection{Trust and Privacy}
Many participants described the typical level of trust placed in a vent channel as strange and believed it to be undeserved. The type of content shared with a vent channel is frequently sensitive and personal information, and offline, this information is usually shared only with close and trusted individuals \cite{Greene2006-pl}. Many vent channels are not restricted, meaning that the entire server is able to view and send messages in the channel. In theory, in most servers, it is very easy for any sort of bad actor to infiltrate a server and its vent channel, which P4 describes has led to harassment of users based on their personal and vulnerable vents.  

In other cases, the number of people in a vent channel could be a positive, and a reason for participating in it. When a vent channel has many users, there are more potential responders to engage with a venter. P1 believes that people may use a vent channel because of the absence of privacy, stating, “I think to a certain extent, people are like, ‘Oh, this is going to reach a lot of people. That's why I'm here. Because, chances are one of them's gonna respond.’”  

\subsection{Vent Channel Features}
How do vent channels differ in different servers? What features of the server affect the vent channel, and in what ways? Participants describe servers that are large (above a few hundred members) or publicly discoverable as being worse for vent channels for multiple reasons. They believe that larger servers have a higher likelihood of interpersonal conflict because of the increased number of members. Additionally, larger servers may have fewer moderators compared to the amount of content posted, which, combined with discoverability, could result in more opportunity for infiltration of a server, harassment, and general bad actors. Another major point that participants describe is a decrease in privacy with an increase in unknown users. P6 is a member of multiple servers with vent channels, but they only use vent channels in smaller servers. They felt uncomfortable posting in a vent channel within a large server, describing, “[there’s] tons and tons of people. And who knows, I have no idea who would be seeing it.” Additionally, large servers tend to be very busy. With an increase in members, an increase in content posted can be reasonably expected. P13 stated, “Having a vent channel for a larger community is like showing up to a group therapy session with like 400 people in it. It probably won't be very productive, and even if it is, it'll probably get drowned out.” 

Some participants believe that servers with a pre-existing supportive environment have better, or more supportive, vent channels. P2 also described effective moderation as essential to a vent channel, saying, 

\begin{quote} I think the most essential thing is just having a kind, understanding community. And also having people to police those channels as well to make sure it's a safe space, because if it's not safe, then no one's really going to feel comfortable enough to speak about their problems or whatever they're struggling with. \end{quote}

Others described multiple other features that they feel improved the vent channel. Some channels had a resource with a list of mental health hotlines “pinned” in the channel, which means that they were displayed in a manner that is easily accessible from the channel itself. Some servers use a Discord bot in order to delete all messages posted in the vent channel on a regular time frame which ranges from days to months. P1 described this as very helpful:  

\begin{quote} If you walk into a discord channel and complain, I think it's a good thing if eventually that goes away and it disappears ... because then you can complain about something without worrying that someone's going to look at it 40 years from now and say, ‘Wow, I can't believe you said that.’ \end{quote}

 One server studied has an anonymous venting feature, in which a user can utilize a bot as a proxy to send a venting message, so that the message is not linked to their Discord account at all. The latter two features can reduce the possibility of negative interaction by limiting the connection between a user and their vents.  

Another feature that Discord has is the ability to limit a channel to users with a specific “role” or tag, which is a method of categorizing users. For example, a user with a moderator role may have the ability to delete messages or view a moderators-only channel, where a general user without that role would not. In a similar way, the vent channel can be restricted to users with a specific role, whether that role is users who opt in to view vent content, users who have posted a certain number of times, or users who have been specifically approved by moderators. This system limits the public accessibility of vent channels, which could help to increase the privacy and decrease vulnerability of individuals who use it.  

\subsection{Communication, Clarity, and Guidelines}
As previously mentioned, many users have conflicting ideas of what a vent channel is and how it should be used, which can lead to miscommunication when there are no explicit rules or guidelines on interaction. Some users may believe that anything posted is fair game to respond to, because Discord is a messaging app built for the purpose of fostering interaction, and typical use of the platform involves communicating with and responding to others. Other users may see a vent channel as a place for venters to be heard first and helped later. Some users might even think of vent channels as a place for frustrations to be virtually shouted into the void and never seen by another user. Users make implicit assumptions as to how the space should operate, assumptions which P4 described as creating a “conflict of interest between the sorts of engagements people are looking for.”  

P5 is a moderator of a Discord server with a vent channel, which used to have no real rules. Because it was a smaller, private, server, he did not have to deal with bad actors, but he did have some issues which he described as “people that were … stepping out of line of boundaries that hadn't really been said.” He and other moderators then revamped the vent channel, as well as other elements of the server, based on previously expressed concerns, and created a “common sense guide of do's and don'ts.” This change included encouraging people to hide content that was likely to be triggering. The "spoiler" feature of Discord was created to prevent people from seeing the ending of media they have not yet watched, and hides text until you click on it \cite{noauthor_2022-st}. The same mechanism can be used to mask sensitive content. The mods also prohibited discussion of violence towards others, and discouraged unsolicited advice. Since the change, he stated that the channel has become easier to engage with, more structured, and less draining. 

In Seering, Kraut, and Dabbish's paper on Twitch moderation, they examine a combination of preventative moderation, such as restricting chat modes, and reactive moderation tools, which enforce punitive actions such as bans after individuals have broken rules \cite{Seering2017-ab}. Seering et. al find that these moderation actions can work in tandem to encourage pro-social behaviors in online chat spaces. P5's experiences align with these results: preventative moderation, such as the use of spoilers, in combination with clearly defined rules and reactive moderation after those rules are broken, promotes pro-social behavior in participants. 

\section{Discussion}
Overall, many participants described having positive experiences with vent channels when they use them to seek social support and subsequently receive it. However, there were many occasions when vent channels can lead to negative experiences for those who use them or read them. Considering this potential for harm, we will discuss some ways to enhance the best functionalities of vent channels and minimize negative situations and experiences.  

Multiple participants, especially those who were moderators for vent channels, mentioned a desire for participants to be informed that a vent channel is not a replacement for professional mental health care such as therapy or psychiatry. When participants are not aware of this discrepancy, they may expect more than can reasonably be provided by other peers, and they may subsequently be disappointed with their experiences in the channel. A good way to improve the effect of vent channels and improve the experience of those who use them could be to align user expectations with what the channel can reasonably provide. Generally, as described by participants, a vent channel can provide a place for people to be heard and acknowledged, especially in small ways such as an emoji reaction. A vent channel can provide peers to talk to, although they are not professionally trained and could in some instances make the situation worse. A vent channel has the unique positive of generally providing validation and can give venters access to people with similar difficulties, such as when members of marginalized communities vent about their experiences as a minority. However, a vent channel cannot provide in-depth talk therapy with the necessary cognitive work for fully resolving frustrations and emotional issues, nor is it typically able to help people identify and resolve concerning thoughts and behaviors to a significant degree.  

In light of this limitation, there are some ways that vent channels could provide more transparency to participants about what they can offer. Vent channel moderators could provide very visible disclosures when first accessing the channel that state that the channel is not intended to serve as a replacement for professional healthcare but as a place to express frustrations or sadness at difficult emotions or situations (and, if relevant to the particular server, a place to receive support from other members). In cases of apparent crisis where a member expresses a problem beyond the ability of the vent channel, moderators could refer these members to other organizations such as Crisis Text Line, which operates in a similar chat-based format but use volunteers who have received more training \cite{noauthor_2021-wi, Yao2022-rp}.  

Additionally, vent channels should be set up to provide structure, clarity, and transparency to members whenever possible. Participants should be aware of the norms of the space before interacting: is this channel purely a place to vent where content is not engaged with at all? Do people provide low-effort support through reactions, or do they engage with the venter through detailed messages? Is the expectation for responders to always provide validation, or do they offer gentle suggestions and critique as well? Venters should be encouraged to specify the type of interaction they desire after posting so that they can receive it. For example, "I'm not looking for advice, I want to know if I'm reasonable or not," "Could I get some support? I want to know if things will be ok." If participants know what to expect, they are less likely to be disappointed or emotionally hurt, and more likely to have a positive experience with the vent channel.

Another common concern of participants was the overwhelming nature of a consistently negative vent channel. Reading a channel designed to be a home for negative feelings could have an effect similar to “doomscrolling,” where participants feel emotionally drained and sad that others are suffering. To help combat this effect, vent channels could widely implement the practice of preventing content from being automatically displayed, using a feature called “spoilering.” When text is spoilered, it is blacked out and not visible to a reader until they choose to see it and “despoiler” by tapping or clicking on the text. This change gives readers more of a choice when deciding whether or not to interact with sensitive and potentially draining content. Participants described using pre-existing mechanisms for this purpose, such as using Discord’s built-in not safe for work (NSFW) channel type so that messages sent in the vent channel don't display previews in notifications or the in-app inbox. This solution helps to keep the vent channel integrated into the server while increasing user agency.

Providing participants with increased clarity about the ability of the vent channel can enable them to seek the appropriate level of social support, rather than expecting support on the level of professional help. Additionally, by reducing the potential for harm in vent channels and making them safer spaces, we can enable participants to achieve perceived connectedness and authenticity with others.

In terms of design recommendations, many of the above features could be either implemented into Discord’s default settings or recommended with set-up instructions in a moderator guide for vent channels. Discord could implement a new channel type, similar to the existing categories of general text channels, announcement channels, rules channels, forum channels, and NSFW channels. This channel could be given a new icon to visually distinguish it from the others and have automatically suggested guidelines and links to Discord-partnered external resources. It could also configure role permissions such that only people with an opt-in vent role could read channel messages and some sort of feature to automatically blackout or “spoiler” text that discusses sensitive content. For example, NSFW channels on Discord don’t display message previews in notifications or in the inbox, and vent channels could have a similar restriction. 

\section{Limitations and Future Work}
This paper was conducted from a qualitative perspective, so we have no information on how common the discussed themes are and how well they represent the average experience with vent channels. Additionally, due to the closed nature and limited discoverability of the majority of servers on Discord, participants in this study were recruited via convenience sampling, and our results cannot be generalized to the broader population of vent channel users. Other limitations include the fact that participants were not provided compensation for their time interviewing, which could have influenced the available sample size.  

A final and important limitation is that we were unable to interview minors due to the difficulty of obtaining parental consent. While conducting preliminary research, the first author observed multiple servers centered around venting with an audience that was majority underage, based on self-assigned user roles. One participant also stated that the server he moderates is majority-underage, and that this demographic skew greatly influences the vent channels within his server. By limiting our work to adults, we are unable to understand some aspects of vent channels and some cases when they could become severely harmful, especially for a vulnerable audience.  

In future work, quantitative analysis would complement our qualitative findings. For example, one might use sentiment analysis before and after the addition of a vent channel to a server. Interviewing minors to gain understanding of their experiences with vent channels would be beneficial, if a way could be found to do so ethically without compromising their privacy (including privacy from their parents). 

\section{Conclusion}
In the movie \textit{Lost in Translation}, two strangers (played by Bill Murray and Scarlett Johansson) sit down at a bar and share their worries. Being in a bar far from home gives them a context to share their life situation with someone they have never met before.
On Discord vent channels, that stranger willing to listen is always available.
Vent channels are a prevalent phenomenon in Discord servers, even those not specifically focused on mental health. Many users have positive experiences with them and are able to increase their well-being via the four mechanisms described in Luo and Hancock's framework \cite{Luo2020-sp}. Vent channels can provide connectedness with others, social support, and the ability for participants to express themselves authentically. They can help venters to feel comforted and supported, and they can serve as a space to receive advice about difficult situations. At the same time, vent channels can cause negative experiences, either due to the emotional intensity of content posted there, fatigue from extensively supporting others, or unclear expectations about how interactions in the vent channel should occur. In addition, the topics discussed in vent channels can be more intensive than may be fit for an informal environment, as the vent channels observed included discussion of sensitive topics such as suicidality and self-harm. Many users had experiences with being concerned for the safety of another user due to the content that they vented about. To improve positive experiences and decrease the potential for negative experiences, we suggest that moderators of vent channels should be encouraged to provide clear guidelines to users about what they can reasonably expect from a vent channel, and that venters should be explicit to potential responders about the kind of response that they would like to receive. 

\bibliographystyle{acm}
\bibliography{references}

\newpage

\section{Appendix}
\subsection{Interview Questions}

\label{interview-questions}
\renewcommand{\labelenumi}{\arabic{enumi}.}
\renewcommand{\labelenumii}{\alph{enumii}.}
\begin{enumerate}
    \item Introductory Questions/Broader Server Experience 
    \begin{enumerate}
        \item How long have you used Discord?  
        \item What sorts of communities did you join on Discord initially, and which communities do you use Discord for now? 
        \item What first drew you to this server?  
        \item Tell me about this server and your experience with it. 
        \item Are you a part of multiple servers that have vent channels?  
        \item Could you tell me briefly about those servers and their vent channels? 
        \item Could you compare and contrast for me the vent channels in these different servers? 
    \end{enumerate}
    \item Experience with Venting
    \begin{enumerate}
        \item Why did you choose to engage with and interact with this particular server?  
        \item How often do you participate in the vent channel, and what do you typically do when you connect? 
        \item Tell me about a time you vented on this channel. What did you share, and how did others respond?
        \item Have you ever vented and received no response? How did that make you feel? 
        \item When you vent, what are you looking for? Would you like to receive support, do you just want to get your thoughts out ("scream into the void"), or is it neither of those?
        \item Do you ever worry about your privacy when you share things on the channel? 
    \end{enumerate}
    \item Experience as a Supporter/Observer  
    \begin{enumerate}
        \item Tell me about a time you responded to someone on the vent channel. 
        \item How often do you spend time on the channel, and what do you typically do? 
        \item What is your relationship like to people on the channel? 
        \item How do you feel after you’ve spent time on the channel?
        \item Has anyone ever worried you because of what they posted on the channel? What did they say and what happened as a result? 
        \item Why do you think people use the channel? 
    \end{enumerate}
    \item Experience as a Moderator 
    \begin{enumerate}
        \item How often do you spend time viewing the vent channel? 
        \item Do you moderate content in the vent channel? If so, what sorts of things do you moderate?  
        \item How do you feel about the vent channel?  
        \item Were you involved in the decision to create a vent channel? If so, how did you, or the broader moderation team, make that decision? 
        \item Does the vent channel have rules? If so, how were those crafted? Have you had instances of rule violation? If so, could you tell me about them?  
    \end{enumerate}
    \item Other/Miscellaneous
    \begin{enumerate}
        \item What features of a vent channel are essential and help it function?  
        \item How would you describe the purpose of a vent channel? 
        \item What interesting rules or features have you seen in vent channels? What effects have you seen those had on interactions in the channel? 
        \item Have you seen any happy-venting channels? What relation do you think those have to primary vent channels? 
        \item Is there anything else that you’d like to tell me?  
    \end{enumerate}
    \item Demographics
    \begin{enumerate}
        \item How old are you?  
        \item What is your gender? 
        \item What are your pronouns?
        \item What is your ethnicity?  
        \item What country do you live in? 
    \end{enumerate}
\end{enumerate}

\end{document}